\documentclass[final,1p,times,twocolumn]{elsarticle}
\usepackage[T1]{fontenc}
\usepackage{epsfig}
\usepackage{amsmath}
\usepackage{amssymb}
\usepackage{color}
\usepackage{dsfont}
\usepackage{epstopdf}

\newcommand{\p}{\partial}
\DeclareMathAlphabet{\bi}{OML}{cmm}{b}{it}

\newcommand{\bsigma}{\boldsymbol{\sigma}}
\newcommand{\bdel}{\boldsymbol{\delta}}
\newcommand{\bnab}{\boldsymbol{\nabla}}

\journal{Solid state communication}

\begin{document}
\begin{frontmatter}
\title{Pseudo magnetic field in strained graphene: revisited}
\author{M. Ramezani Masir}\
\author{D. Moldovan}
\author{F. M. Peeters}
\address{Departement Fysica, Universiteit Antwerpen \\
Groenenborgerlaan 171, B-2020 Antwerpen, Belgium.}
\begin{abstract}
We revisit the theory of the pseudo magnetic field as induced by
strain in graphene using the tight-binding approach. A systematic
expansion of the hopping parameter and the deformation of the
lattice vectors is presented from which we obtain an expression for
the pseudo magnetic field for low energy electrons. We generalize
and discuss previous results and propose a novel effective
Hamiltonian. The contributions of the different terms to the pseudo
magnetic field expression is investigated for a model triaxial
strain profile and are compared with the full solution. Our work
suggests that the previous proposed pseudo magnetic field expression
is valid up to reasonably high strain ($ 15\%$) and there is no
${\bf K}$-dependent pseudo-magnetic field.
\end{abstract}
\end{frontmatter}
\section{Introduction}
Graphene has triggered a broad activity both in fundamental and
applied physics and chemistry. The most intriguing feature of this
system is their similarity to ultrarelativistic electrons and
positrons obeying the Dirac equation \cite{no,zh}. An interesting
prediction is that a geometrical deformation of the graphene lattice
results in a local strain that acts as a pseudo-magnetic field on
the electronic degrees of freedom and which leads to a
pseudo-quantum Hall effect \cite{M1}. Graphene can sustain very
high, up to $25\%$, elastic strains \cite{S0} which leads to a shift
in the position of the Dirac cones \cite{M11}. Deformation due to
elastic strain changes the hopping amplitude of the carbon atoms and
induces an effective vector potential that shifts the Dirac point
\cite{M5}. With a proper geometrical deformation it is possible to
create large pseudo-magnetic fields of different shapes \cite{M1,
M6,Z1}. It has been predicted that applying strain with triangular
symmetry results in an uniform pseudo-magnetic field of the order of
$~10$T \cite{M4}. Recently it was reported experimentally \cite{M12}
that nanobubbles grown on a Pt(111) surface induce pseudo-magnetic
fields of more than $300$ T. Landau quantization of the electronic
spectrum was observed by scanning tunneling microscopy. Thus strain
engineering has become a new way to control the electronic
properties of graphene \cite{Z1,Z3}.

The effective vector potential induced by strain was derived in
Refs. \cite{X0,X1} and was based on a tight binding approach with
the important approximation that the local strain does not alter the
lattice vectors. Very recently it was shown that including the
deformation of the lattice vectors leads to an extra term for the
effective magnetic field which is of the same order of magnitude and
which differ in the different $K$ points \cite{X2}. But later it was
shown that this extra term in the effective vector potential does
not have any contribution to the induced pseudo magnetic field
\cite{X22} and that subsequently there is no different in the
different ${\bf K}$-points. Furthermore, in Ref. \cite{X3,X4} it is
shown that in the presence of strain the Fermi velocity becomes
spatial dependent.

In this manuscript we revisit the problem and present a systematic
study of the different corrections to the vector potential and
compare them with the numerically obtained full pseudo magnetic
field. We present the effective Hamiltonian that includes different
contributions of strain. The previous result for the vector
potential and the Fermi velocity reobtained in our systematic
expansion. As an example we present explicit analytical results for
strained graphene as induced by a uniaxial and triaxial strain. We
find the magnetic field induced by the in-plane deformation and
compared the different terms for the vector potential
\cite{X0,X1,X2} with the exact numerical results for the pseudo
magnetic field.
\section{Strain Field}
\label{sec1}
The Hamiltonian in the tight-binding approximation
considering only the first nearest neighbor is given by:
\begin{equation}\label{H0}
    \displaystyle{H = -\sum_{i,j}t_0 a_{i}^{\dag}b_{j} + h.c.},
\end{equation}
where $t_0$ is the hopping parameter and $a_{i}$ and $a^{\dag}_{i}$ ($b_{i}$
and $b^{\dag}_{i}$) are the annihilation and creation operator for
an electron on sublattice A (B). In the presence of lattice
deformation the hopping parameter $t$ changes due to the changing
interatomic distance. The modification of the hopping parameter
due to strain is given by \cite{M11},
\begin{equation}\label{hopping}
\displaystyle{t_{n} = t_0 e^{-\beta(d_{n}/a - 1)}},
\end{equation}
\begin{figure}[ht]
  \begin{center}
  \includegraphics[width=8cm]{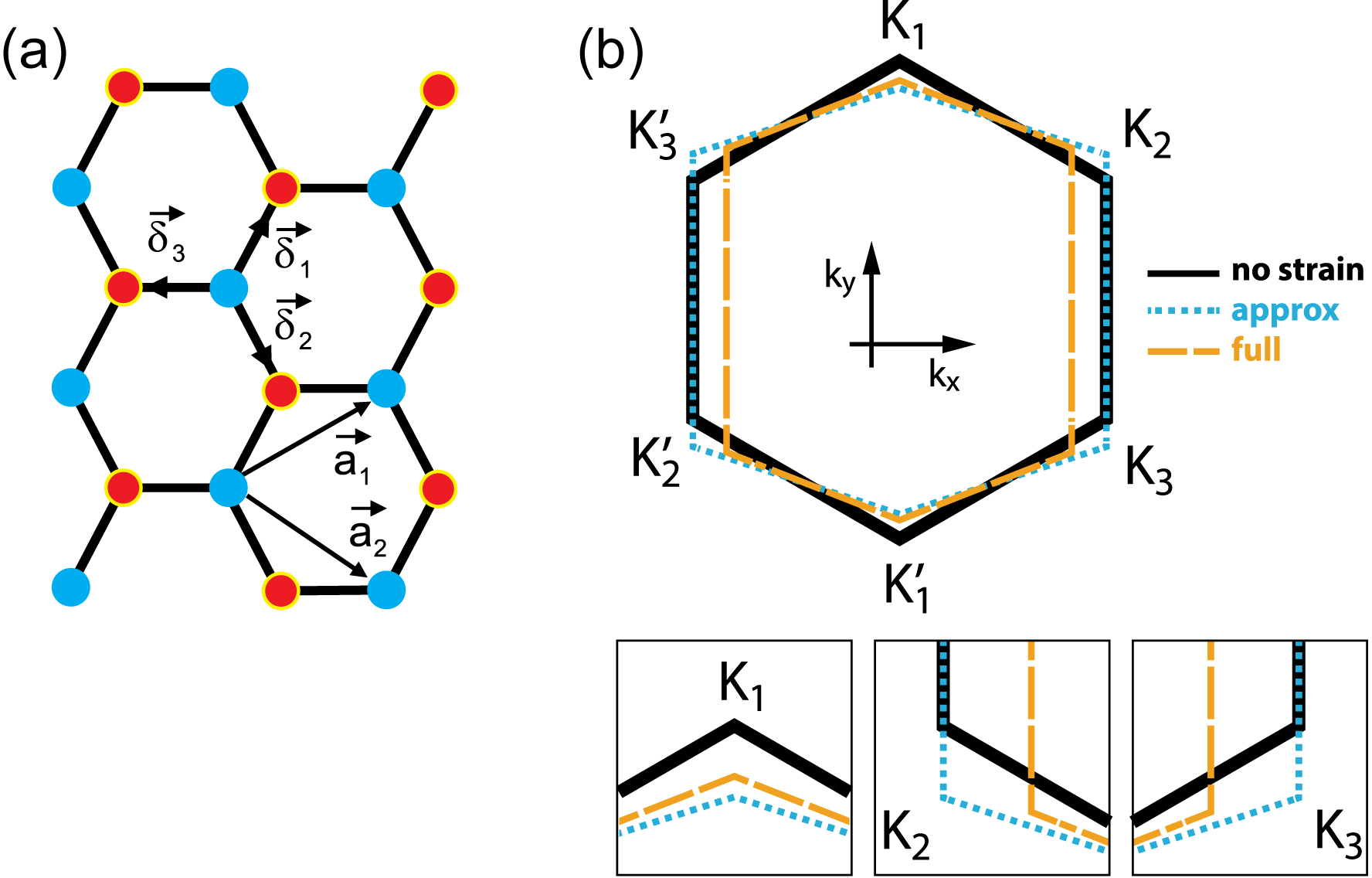}
   \caption{(a) A monolayer graphene lattice, ${\bf {a}}_{1} =
   (3/2,\sqrt{3}/2)a$ and ${\bf{a}}_{2} = (3/2,-\sqrt{3}/2)a$
   are the basis vectors and the sublattices are connected by
   ${\boldsymbol{\delta}}_{1} = (1/2,\sqrt{3}/2)a$,
   ${\boldsymbol{\delta}}_{2} = (1/2,-\sqrt{3}/2)a$ and
   ${\boldsymbol{\delta}}_{3} = (-1,0)a$. (b) The normal
   (solid black line) and deformed (yellow dashed line for
    full solution and blue dotted for approximated pseudo
    magnetic field) Brillouin zone with ${\bf K}$ points given
    by ${{\bf{K}_{1}} = \frac{4\pi}{3\sqrt{3}a}\left(0,1\right)}$,
     ${{\bf K}_2 = \frac{2\pi}{3\sqrt{3}a}\left(\sqrt{3},-1\right)}$
     and ${{\bf K}_3 = \frac{2\pi}{3\sqrt{3}a}\left(-\sqrt{3},-1\right)}$ .} \label{F0}
  \end{center}
\end{figure}
where $a$ is the unstrained nearest neighbor distance, $\beta
\approx 2 - 3.37$ and $d_{n}$ is the length of the strained lattice
vector. Using the Fourier transform of the creation and annihilation
operators we obtain the strained Hamiltonian as:
\begin{equation}\label{HS}
    \displaystyle{H = -\sum_{n,{\bf k}} t_n e^{- i{\bf k}\cdot {\bf d}_{n}}a_{\bf k}^{\dag}b_{\bf k}
     + H.c.}
\end{equation}
${\bf{d}_n} = (\bar{I} + \bar{u}){\bdel_{n}},$ where $\bar{I}$ is
the unity matrix and $\bar{u}$ is the strain tensor. The strain
elements of the tensor are given by \cite{X4,X22}, $\bar{u} =
\bar{\epsilon} + \bar{\omega}$, and it consists of two parts: the
linear part of strain tensor given by,
\begin{equation}\label{GA2}
    \displaystyle{\bar{\epsilon}_{i j} = \frac{1}{2}\left\{\frac{\p u_{i}}{\p x_{j}}
    +\frac{\p u_{j}}{\p x_{i}} \right\}}.
\end{equation}
and the rotational part $\bar{\omega}$ given by
\begin{equation}\label{GA2}
    \displaystyle{\bar{\omega}_{i j} = \frac{1}{2}\left\{\frac{\p u_{i}}{\p x_{j}}
    -\frac{\p u_{j}}{\p x_{i}} \right\}}.
\end{equation}
On the other hand we can obtain the change in the lattice vectors size as
\begin{equation}\label{E11}
\begin{split}
    d \ell'^2& = dx^{'}_{i} dx'_{i} = dx_{i}dx^{i} + 2du_{i}dx_{i} + du_{i}du_{i}\\
    & =d \ell^2 + 2\frac{du_{i}}{dx_{k}}dx_{i}dx_{k} +
    \frac{du_{l}}{dx_{k}}\frac{du_{l}}{dx_{k}}dx_{i}dx_{k}\\
    & = d \ell^2 + 2\left\{\underbrace{\frac{1}{2}\left(\frac{du_{i}}{dx_{k}}+ \frac{du_{k}}{dx_{i}}\right)dx_{i}dx_{k}}_{1}
    +\underbrace{\frac{1}{2}\left(\frac{du_{i}}{dx_{k}} - \frac{du_{k}}{dx_{i}}\right)dx_{i}dx_{k}}_{2}\right\}+ \frac{du_{l}}{dx_{k}}\frac{du_{l}}{dx_{k}}dx_{i}dx_{k}\\
    & = d \ell^2 + 2\left\{\bar{\epsilon}_{ij} dx_{i}dx_{k}
    + \bar{\omega}_{ij} dx_{i}dx_{k}\right\}+ \frac{du_{l}}{dx_{k}}\frac{du_{l}}{dx_{k}}dx_{i}dx_{k}.\\
\end{split}
\end{equation}
The first symmetric term denoted by $1$ correspond with the linear
part of the strain and the second term denoted by $2$ correspond
with rotational tensor and has zero contribution to the nearest
neighbor vector sizes. The hopping changes with carbon-carbon
distance but the rotational tensor term does not contribute to it.

First we drive the effective Hamiltonian by expanding  Eq.
(\ref{HS}) up to the first order in strain \cite{X2} (The second
order terms are included in the subsequent discussion but are not
listed in the expansion of the Hamiltonian because of those
expressions are rather involved),
\begin{equation}\label{Hamil0}
\begin{split}
    H &= -\sum_{n=1}^{3} t_{n}
    \left(
      \begin{array}{cc}
        0 & e^{-i({\bf K} + {\bf q})\cdot {\bf d_{n}}} \\
        e^{i({\bf K} + {\bf q})\cdot {\bf d_{n}}} & 0 \\
      \end{array}
    \right)\\
    &\approx -\sum_{n=1}^{3} t_{n}
    \left(
      \begin{array}{cc}
        0 & e^{-i{\bf K}\cdot{\bf a_{n}}} \\
        e^{i{\bf K}\cdot{\bf a_{n}}}& 0 \\
      \end{array}
    \right)(1 + i \sigma_{z}{\bf K}\cdot \bar{u}{\bf a}_{n}) (1 + i\sigma_{z}{\bf q}\cdot{\bf d}_{n})\\
    & = -\sum_{n=1}^{3} t_{0}\left(1 - \frac{\beta}{a^2}{\bf a}_{n}\cdot \bar{u}\cdot {\bf a}_{n}\right)
    \left(\frac{i}{a}(\bsigma \cdot {\bf a}_{n})\sigma_{z}\right)(1 + i \sigma_{z}{\bf K}\cdot \bar{u}
    {\bf a}_{n})\\
    &~~~~~~~~~~~~~~~~~~~~~~~~~\times (1 + i\sigma_{z}{\bf q}\cdot{\bf a}_{n} + i\sigma_{z}{\bf q}
    \cdot\bar{u}{\bf a}_{n}).\\
\end{split}
\end{equation}
The different terms of the effective Hamiltonian are shown in Table. 1.
The first term is the famous Dirac-Weyl equation,
\begin{equation}\label{Hamil0}
\begin{array}{l}
  \displaystyle{H_{0} = -\sum_{n=1}^{3} t_{0} \left(\frac{i}{a}(\bsigma \cdot {\bf a}_{n})\sigma_{z}
  \right)\left(i\sigma_{z}{\bf q}\cdot{\bf a}_{n}\right) = v_{0} \bsigma \cdot {\bf p}},\\
  \displaystyle{H_{0}  = -i \hbar v_{0} \bsigma \cdot {\bnab}}
\end{array}
\end{equation}
There are three terms in the first order of strain,
\begin{equation}\label{Hamil1}
\begin{array}{l}
  \displaystyle{H_{1} = \sum_{n=1}^{3} t_{0} \left(\frac{1}{a}(\bsigma \cdot {\bf a}_{n})\sigma_{z}\right)\left(\sigma_{z}{\bf q}\cdot\bar{u}{\bf a}_{n}\right) = v_{0} \bsigma \cdot \bar{u} \cdot {\bf p}},\\
  \displaystyle{H_{1}  =  -i \hbar v_{0}\left( \bsigma \cdot \bar{u}\cdot {\bnab} + \frac{1}{2}\bsigma \cdot {\bnab}^{T}\cdot \bar{u}\right)}
\end{array}
\end{equation}
the second term is $\beta$-dependent and is given by
\begin{equation}\label{Hamil2}
\begin{array}{l}
  \displaystyle{H_{2} = \sum_{n=1}^{3} t_{0} \left(\frac{\beta}{a^2}{\bf a}_{n}\cdot \bar{u}
  \cdot {\bf a}_{n}\right)\left(\frac{1}{a}(\bsigma \cdot {\bf a}_{n})\sigma_{z}\right)
  \left(\sigma_{z}{\bf q}\cdot{\bf a}_{n}\right)}\\
  ~~~~=\displaystyle{ \frac{\beta v_{F}}{4}  \bsigma \cdot
  (2\bar{u} + Tr(u)\bar{I}) \cdot {\bf p}},\\
  \\
  \displaystyle{H_{2}  =  -i \hbar \frac{\beta v_{F}}{4}  \bsigma \cdot \left(2\bar{u} \cdot
  \bnab + Tr(u)\bar{I}\cdot \bnab + \bnab^{T} \cdot \bar{u} + \frac{1}{2}\bnab^{T} \cdot
  Tr(u)\bar{I}\right)}
\end{array}
\end{equation}
and is the same as the term introduced in Refs. \cite{X3,X4}. The
third and last term is $\beta$-independent and is given by
\begin{equation}\label{Hamil3}
\begin{array}{l}
  \displaystyle{H_{3} = -\sum_{n=1}^{3} t_{0} \left(\frac{i}{a}(\bsigma \cdot {\bf a}_{n})\sigma_{z}
  \right) \left( i\sigma_{z}{\bf K}\cdot \bar{u}{\bf a}_{n}\right)\left(i\sigma_{z}{\bf q}
  \cdot{\bf a}_{n}\right)}\\
  \displaystyle{~~~~= i\frac{v_{F}^{0}a}{2}\bsigma\cdot ({\bf K} \cdot \bar{u} \cdot \boldsymbol{\omega} )\cdot {\bf p}}\\
  \displaystyle{H_{3}= \hbar \frac{v_{F}^{0}a}{2} \bsigma\cdot \left\{({\bf K} \cdot \bar{u} \cdot \boldsymbol{\omega} )
  \cdot {\bnab} + \frac{1}{2}{\bnab}\cdot({\bf K} \cdot \bar{u} \cdot \boldsymbol{\omega} )\right\}}\\
\end{array}
\end{equation}
Here $v_{F}^{0} = 3ta/2\hbar$, $\bsigma = (\sigma_{x}, \sigma_{y})$ are the Pauli matrices, and $\boldsymbol{\omega}=
(-\sigma_{z},\sigma_{x})$. In summary we can write
the full effective Hamiltonian up to the first order in strain, considering both the
$\beta$-dependent and $\beta$-independent terms as
\begin{equation}\label{HAmfull}
\begin{split}
    H_{eff} &= \displaystyle{H_{0} + H_{1} + H_{2} + H_{3}}.
\end{split}
\end{equation}
Now the $\beta$-dependent Fermi velocity is replaced by a tensor,
%
\begin{equation}\label{vf}
    \displaystyle{\bar{v}_{F} = v_{F}^{0}\left(\bar{I} +  \frac{\beta}{4} [2\bar{u}
   + Tr(u)\bar{I}] \right)}
\end{equation}
which is space-dependent \cite{X3,X4}.
%

Next we derive the pseudo magnetic field induced by strain. The
pseudo-magnetic vector potential $A_{ps} = A_x + i A_{y}$ induced by
strain is given by \cite{M11},
\begin{equation} \label{Atot}
  A_{ps} = \frac{1}{e v_F} \sum_n^3 t_n e^{-i {\bf K} \cdot {\bf d}_n},
\end{equation}
where $v_F$ is the Fermi velocity, $t_n$ are the strained
nearest-neighbor hopping parameters. Note that $A_{ps}$ is imaginary
because strain breaks inversion symmetry in the nearest neighbor
hopping. The effective pseudo magnetic field induced by strain will
shift the ${\bf K}$-points as ${\bf K}_{n} \to {\bf K}_{n} + {\bf
A}_{n}$ (see Fig. \ref{F0}(b)). Writing the wave vector ${\bf k}$
with respect to the Dirac cone using ${\bf k} = {\bf K} + {\bf q}$
and expanding the exponent and hopping parameter $t_{n}$ up to
second order we find:
\begin{equation}\label{vf1}
\begin{split}
    t_{0}\left(1 + \delta t_{n} + \frac{1}{2}\delta t^2_{n}\right)&\left(1 - i {\bf K}\cdot
    \bar{u}\cdot {\bdel}_{n} -  \frac{1}{2}({\bf K}\cdot
    \bar{u}\cdot {\bdel}_{n})^2 \right)\\
    &\times(1 - i {\bf q}\cdot{\bdel}_{n} - i {\bf q}\cdot
    \bar{u}\cdot{\bdel}_{n})e^{-i {\bf K}\cdot {\bdel}_{n}}
\end{split}
\end{equation}
where $\delta t =
-\frac{\beta}{a^2}\bdel_{n}\cdot\bar{u}\cdot\bdel_{n}$.
\begin{table*}[!]
\caption {Different terms induced by strain in the expansion of the vector potential.
Right column indicates the order of these terms in the strain (i. e. $O(u^2)$) and their effect on the different
properties.} \label{table1}
\begin{center}
    \begin{tabular}{ | l |l| p{6.5cm} |}
    \hline
    n&Expansion terms  &  \\ \hline
    1 &$e^{-i {\bf K}\cdot { \bdel}_{n}}$  & ~~0 \\ \hline
    2 &$-i {\bf q}\cdot{ \bdel}_{n}e^{-i {\bf K}\cdot { \bdel}_{n}}$  & ~~Dirac equation \\ \hline
    3 &$-i {\bf q}\cdot\bar{u}\cdot{ \bdel}_{n}e^{-i {\bf K}\cdot { \bdel}_{n}}$ & ~~$v_{F}$ $\beta$-independent \\ \hline
    4 &$-\frac{\beta}{a^2}\left({ \bdel}_{n}\cdot
    \bar{u}\cdot{ \bdel}_{n}\right)e^{-i {\bf K}\cdot { \bdel}_{n}}$ & Effective vector potential $A_{0}$\\ \hline
    5 &$\frac{i\beta}{a^2}\left({ \bdel}_{n}\cdot
    \bar{u}\cdot{ \bdel}_{n}\right)\left({\bf q}\cdot
    { \bf a}_{n}\right)e^{-i {\bf K}\cdot { \bdel}_{n}}$& Fermi velocity $v_{F}$ $\beta$-dependent ~$O(1)$\\ \hline
    6 &$\frac{i\beta}{a^2}\left({ \bdel}_{n}\cdot
    \bar{u}\cdot{ \bdel}_{n}\right)\left({\bf q}\cdot
    \bar{u}\cdot{ \bdel}_{n}\right)e^{-i {\bf K}\cdot { \bdel}_{n}}$& Fermi velocity $v_{F}$ $\beta$-dependent ~$O(2)$\\ \hline
    7 & -$i\left({ \bf K}\cdot
    \bar{u}\cdot{ \bdel}_{n}\right)e^{-i {\bf K}\cdot { \bdel}_{n}}$& Effective ${\bf A}$~ $\beta$-independent $O(1)$\\ \hline
    8 & $\frac{i\beta}{a^2}\left(\bdel_{n}\cdot
    \bar{u}\cdot{ \bdel}_{n}\right)\left({ \bf K}\cdot
    \bar{u}\cdot{ \bdel}_{n}\right)e^{-i {\bf K}\cdot {\bdel}_{n}}$& Effective ${\bf A}$~ $\beta$-dependent $O(2)$\\ \hline
    9 & $\frac{\beta^2}{2a^4}\left(\bdel_{n}\cdot
    \bar{u}\cdot{ \bdel}_{n}\right)^2e^{-i {\bf K}\cdot { \bdel}_{n}}$& Effective ${\bf A}$~ $\beta$-dependent $O(2)$\\ \hline
    10 & $-  \frac{1}{2}({\bf K}\cdot
    \bar{u}\cdot { \bdel}_{n})^2e^{-i {\bf K}\cdot { \bdel}_{n}}$& Effective ${\bf A}$~ $\beta$-independent $O(2)$\\ \hline
    11 & $-\left({ \bf K}\cdot
    \bar{u}\cdot{ \bdel}_{n}\right)({\bf q}\cdot{ \bdel}_{n})e^{-i {\bf K}\cdot { \bdel}_{n}}$& ~~$v_{F}$ $\beta$-independent $O(1)$\\ \hline
    12 & $-\left({ \bf K}\cdot
    \bar{u}\cdot{ \bdel}_{n}\right)({\bf q}\cdot\bar{u}\cdot{ \bdel}_{n})e^{-i {\bf K}\cdot { \bf
    d}_{n}}$& ~~$v_{F}$ $\beta$-independent $O(2)$\\ \hline
    13 & $\frac{\beta}{a^2}({ \bdel}_{n}\cdot
    \bar{u}\cdot{ \bdel}_{n})\left({ \bf K}\cdot
    \bar{u}\cdot{ \bdel}_{n}\right)({\bf q}\cdot{ \bdel}_{n})e^{-i {\bf K}\cdot { \bdel}_{n}}$& ~~$v_{F}$ $\beta$-dependent $O(2)$\\ \hline
    14 & $-\frac{\beta^2}{2 a^4}({ \bdel}_{n}\cdot
    \bar{u}\cdot{ \bdel}_{n})^2({\bf q}\cdot{ \bdel}_{n})e^{-i {\bf K}\cdot { \bdel}_{n}}$& ~~$v_{F}$ $\beta$-dependent $O(2)$\\ \hline
    \end{tabular}
\end{center}
\end{table*}
The effective vector potential is given by ${\bf q}$ independent
terms. Keeping the hopping parameters up to second order and
expanding $e^{-i{\bf K}\cdot {\bf d}_{n}}$ we find
\begin{equation}\label{sc1}
\begin{split}
    t_{n} e^{-i {\bf{K}}\cdot {\bf{d}}_n} & \approx t_{0}\left\{1 + \underbrace{\delta t_{n} }_{1}
    - \underbrace{i {\bf K}\cdot \bar{u} \cdot \bdel_{n}}_{2}\right. \\
    &-\left. \underbrace{ i \delta t_{n} {\bf K}\cdot \bar{u} \cdot \bdel_{n} -  \frac{1}{2}({\bf K}\cdot
    \bar{u}\cdot {\bdel}_{n})^2+ \frac{1}{2}\delta t_{n}^2}_{3}\right\} e^{-i {\bf K} \cdot \bdel_{n}}
\end{split}
\end{equation}
The first correction term is the one obtained in Refs. \cite{X0,X1} and the second
correction term was recently added by Kit {\it{et al.}} \cite{X2}.
The third term is the new higher order correction term which we will add.

Considering only the first term we take constant lattice vectors
${\bf{d}}_{n} = {\bdel}_{n}$. Using the three nearest neighbors
vectors in real space (as shown in Fig. \ref{F0})
$\displaystyle{{\bdel}_{1} = \frac{a}{2}(1,\sqrt{3})}$,
$\displaystyle{{\bdel}_{2} = \frac{a}{2}(1,-\sqrt{3}) }$,
$\displaystyle{{\bdel}_{3} = a(-1,0) }$ and the position of the
${\bf K}$-points are given by $\displaystyle{{\bf{K}_{1}} =
\frac{4\pi}{3\sqrt{3}a}\left(0,1\right)}$, $\displaystyle{{\bf K}_2
= \frac{2\pi}{3\sqrt{3}a}\left(\sqrt{3},-1\right)}$ and
$\displaystyle{{\bf K}_3 =
\frac{2\pi}{3\sqrt{3}a}\left(-\sqrt{3},-1\right)}$ we obtain the
vector potential in terms of the strain tensor elements,
\begin{equation}\label{Ac1}
\begin{split}
    {\bf A}_{1} =  \frac{\phi_0\beta}{4\pi a}
    \left(
    \begin{array}{c}
    u_{xx} -  u_{yy}\\
    2u_{xy}\\
    \end{array}
    \right)
\end{split}
\end{equation}
where $\phi_{0} = h/e$ is the flux quantum.

In order to obtain the correction given in Ref. \cite{X2} we need to
include the change of the lattice vectors with deformation as
${\bf{d}_n} = (\bar{I} + \bar{u}){\bdel_{n}}$. Including this
correction we find the following extra term to the vector potential
for the different ${\bf K}$-points,
\begin{equation}\label{Cor1}
    \begin{array}{l}
          {\bf A}^{K_1}_{2} = \frac{\phi_{0}}{2a}\frac{4}{3\sqrt{3}}
    \left(
    \begin{array}{c}
    u_{yy} \\
    u_{xy} \\
    \end{array}
    \right), \\
    \\
        {\bf A}^{K_2}_{2} =  \frac{\phi_{0}}{2a}\left(
                                               \begin{array}{c}
                                                 \frac{2}{3}u_{xy} - \frac{2\sqrt{3}}{9}u_{yy} \\
                                                 \frac{2}{3}u_{xx} - \frac{2\sqrt{3}}{9}u_{xy} \\
                                               \end{array}
                                             \right), \\
        \\
    {\bf A}^{K_3}_{2} =  \frac{\phi_{0}}{2a}\left(
                                               \begin{array}{c}
                                                 -\frac{2}{3}u_{xy} - \frac{2\sqrt{3}}{9}u_{yy} \\
                                                 -\frac{2}{3}u_{xx} - \frac{2\sqrt{3}}{9}u_{xy} \\
                                               \end{array}
                                             \right).
    \end{array}
\end{equation}
It is possible to show that this effective vector potential has the
form of $\nabla \chi$. We start with
\begin{equation}\label{Ap2}
\begin{split}
    A_{2} &= -\frac{3t_0 a}{2}\left\{{\bf K}\cdot \bar{u} \cdot {\bf s}\right\}\\
    & = -\frac{3t_0 a}{2}\sum_{i,j}\left\{K_i \bar{u}_{ij}  s_{j}\right\}\\
    & = -\frac{3t_0 a}{2}\sum_{i,j}K_i \left(\frac{\p u_{i}}{\p x_{j}}\right)  s_{j}\\
    & = -\frac{3t_0 a}{2}\bnab({\bf K}\cdot {\bf u})\cdot{\bf s}
\end{split}
\end{equation}
and the two components of the vector potential is given by the real
and complex part of ${\bf A}_2$ as
\begin{equation}\label{Ac1}
    \begin{array}{c}
      A_{x} \propto \p_{x}\left({\bf K}\cdot {\bf u}\right) \\
      A_{y} \propto \p_{y}\left({\bf K}\cdot {\bf u}\right)
    \end{array}
\end{equation}
and the magnetic field is given by ${\bf B}_{2} = \bnab \times {\bf
A}_{2} = 0$, which shows that there is no ${\bf K}$-dependent
pseudo-magnetic fields.
%

Next we include the second order strain part and try to find effective vector
potential,
\begin{equation}\label{th1}
    {\bf A}^{\bf K}_{3} = \underbrace{-i \delta t_{n}{\bf K}\cdot \bar{u} \cdot {\bdel}_{n}}_{I_1}
    + \underbrace{\frac{1}{2}\delta t^2_{n}}_{I_{2}} -  \underbrace{\frac{t_0}{2}({\bf K}\cdot
    \bar{u}\cdot { \bdel}_{n})^2}_{I_3},
\end{equation}
and the corresponding vector potential for the different ${\bf K}$-points is given by
\begin{equation}\label{Cor2}
    \begin{array}{l}
      {I}^{K_1}_{1} =  \frac{\phi_0}{2 a}
    \left(
     \begin{array}{c}
     -\frac{\beta }{3\sqrt{3}}(u_{xx}u_{yy} + 3u_{yy}^2 +2 u_{xy}^2) \\
     \frac{\beta }{\sqrt{3}}(u_{xx}u_{xy} + u_{xy} u_{yy}) \\
     \end{array}
     \right) ,\\
     \\
     {I}^{K_2}_{1} =  \frac{\phi_0}{2 a}\frac{\beta }{3\sqrt{3}}
    \left(
     \begin{array}{c}
     - 2u_{xy}^2 + 3\sqrt{3}u_{xy}u_{yy} + 3\sqrt{3}u_{xx}u_{xy} - 3u_{yy}^2 - u_{xx}u_{yy} \\
     - 3\sqrt{3}u_{xx}^2 + 3u_{xx}u_{xy} - \sqrt{3}u_{yy} u_{xx} - 2\sqrt{3}u_{xy}^2 + 3u_{yy}u_{xy} \\
     \end{array}
     \right),\\
     \\
     {I}^{K_3}_{1} =  \frac{\phi_0}{2 a}\frac{\beta }{3\sqrt{3}}
    \left(
     \begin{array}{c}
     2u_{xy}^2 + 3\sqrt{3}u_{xy}u_{yy} + 3 \sqrt{3} u_{xx} u_{xy} + 3u_{yy}^2 + u_{xx}u_{yy} \\
     - 3\sqrt{3}u_{xx}^2 - 3u_{xx}u_{xy} - \sqrt{3}u_{yy}u_{xx} - 2\sqrt{3}u_{xy}^2 - 3u_{yy}u_{xy} \\
     \end{array}
     \right).
    \end{array}
\end{equation}
The correction corresponding to $\delta t^2/2$ is given by
\begin{equation}\label{dt2}
    I_{2}^{\bf K} = \displaystyle{\frac{1}{2}\delta t^2 \to  \frac{\phi_0}{2 a} \left(
                                   \begin{array}{c}
                                     \frac{\beta^2}{8\pi}(5u_{xx}^2 - 2u_{xx}u_{yy} - 4u_{xy}^2 - 3u_{yy}^2) \\
                                     -\frac{3\beta^2}{2\pi}u_{xy}(u_{xx} + 3u_{yy}) \\
                                   \end{array}
                                 \right)}
\end{equation}
and the vector potential resulting from the last contribution $-  \frac{t_0}{2}({\bf K}\cdot
    \bar{u}\cdot { \bdel}_{n})^2$ in the different ${\bf K}$-points is given by,
\begin{equation}\label{Cor2}
    \begin{array}{l}
      {I}^{K_1}_{3} =  \frac{\phi_0}{3a}\frac{4 \pi}{9}
    \left(
     \begin{array}{c}
     u_{xy}^2 - u_{yy}^2 \\
     -2u_{xy}u_{yy} \\
     \end{array}
     \right) ,\\
     \\
      {I}^{K_2}_{3} =  \frac{\phi_0}{3a}\frac{4 \pi}{9}
    \left(
     \begin{array}{c}
     3u_{xx}^2 - 2u_{xy}^2 - u_{yy}^2 + 2\sqrt{3}u_{xy}(u_{yy} - u_{xx}) \\
     -2(u_{xy} - \sqrt{3}u_{xx})(u_{yy} - \sqrt{3}u_{xy}) \\
     \end{array}
     \right),\\
     \\
       {I}^{K_3}_{3} =  \frac{\phi_0}{3a}\frac{4 \pi}{9}
    \left(
     \begin{array}{c}
     3u_{xx}^2 - 2u_{xy}^2 - u_{yy}^2 + 2\sqrt{3}u_{xy}(u_{xx} - u_{yy}) \\
     -2(u_{xy} + \sqrt{3}u_{xx})(u_{yy} + \sqrt{3}u_{xy}) \\
     \end{array}
     \right).
    \end{array}
\end{equation}
This correction is of second order in the strain and is thus
important for large strains and the corresponding effective field is
position dependent. The most important term is $I_{2} = \frac{1}{2}
\delta t^{2}$ which is ${\bf K}$-independent and it is possible to
show that the two other terms $I_{1}$ and $I_{3}$ have a non-zero
contribution to the vector potential but have zero contribution in
pseudo-magnetic field.
\section{Fermi velocity for uniaxial strain}
The tight-binding Hamiltonian for an infinite sheet of graphene is given by,
\begin{equation}
 H = \begin{pmatrix}
      0&f({\bf k})\\
      f^{*}({\bf k}) & 0\\
     \end{pmatrix}
\end{equation}
where,
\begin{equation}
  f({\bf k}) = \sum_{n=1}^3 t_n e^{i{\bf k}{\bf d_n} }.
\end{equation}
Here, $t_n$ is the strained hopping parameter which is given
by\cite{X3},
\begin{equation}\label{straint}
  t_{n} = t_{0}e^{-\beta \omega_{n}}
\end{equation}
where $\omega_{n} = l_{n}/a_{cc} - 1$. Here $t_0 = -2.8$ eV is the
unstrained hopping parameter, $l_{n}$ is the strained distance to
the nearest neighbor atom $n$, $a_{cc} = 0.142$ nm is the unstrained
carbon-carbon distance and $\beta = 3.37$ is the strained hopping
energy modulation factor. The strained nearest-neighbor vectors are
given by ${\bf d_n} = (1+\bar{u}){\bdel}_{n}$.

\begin{figure}[h!]
  \centering
  \includegraphics[width=8.6cm]{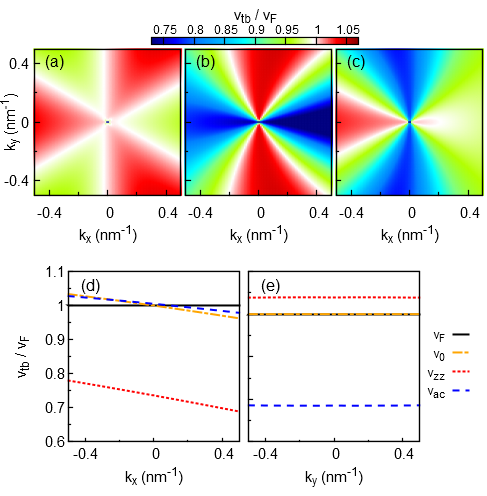}
  \caption{Top: Contour plots of $v_{tb} / v_F$ near the Dirac point for: (a) $v_0$ unstrained graphene,
  (b) $v_{zz}$ uniaxial zigzag strain and (c) $v_{ac}$ uniaxial armchair strain. Bottom: Fermi velocity along
  the cuts where (d) $k_y = 0$ and (e) $k_x = 0$, for all there cases of the velocity: $v_0$, $v_zz$ and
  $v_ac$. The solid black line indicates the traditional continuum limit Fermi velocity $v_F$. The strain
  intensity is $10\%$. }
  \label{fig:vF_uni_1}
\end{figure}

We calculate the energy spectrum $E({\bf k})$ of a graphene sheet
from the tight-binding Hamiltonian. The velocity can then be
obtained as ${\bf v} = {\bf \nabla_k} E({\bf k})$. We calculate the
velocity for three cases: 1) unstrained graphene, 2) graphene
strained in the zigzag (zz) direction and 3) graphene strained in
the armchair (ac) direction. The results are shown in Fig.
\ref{fig:vF_uni_1}. Note that we only consider the part of the
spectrum that is close to the Dirac point where the continuum limit
may be applied (up to $300$ meV). The velocity obtained from
tight-biding is scaled by the traditional continuum limit Fermi
velocity $v_F = \frac{3ta_cc }{2 \hbar}$. In the case of unstrained
graphene from Fig. \ref{fig:vF_uni_1}(a), the deviation of $v_{tb}$
from $v_F$ is generally smaller than $3\%$. However, moving to
strained graphene (see Figs. \ref{fig:vF_uni_1}(b,c)), the velocity
deviates from $v_F$ by as much as $25\%$.
\section{Pseudo-magnetic field for triaxial strain}
The displacement of triaxial strain is given by ${\bf u}({\bf r}) =
(u_x, u_y)$,
\begin{equation}
 \begin{array}{ccc}
  u_{x}&=&2cxy,\\
  u_{y} &=& c (x^2 - y^2),
 \end{array}
\end{equation}
where $c$ is a constant. The corresponding strain tensor
$u_{ij}({\bf r}) = \p_j u_i$ is,
\begin{equation}
 \bar{u}({\bf r}) = c \begin{pmatrix}
                         y & x \\
                         x & -y \\
                     \end{pmatrix}.
\end{equation}

\begin{figure}[h!]
  \centering
  \includegraphics[width=8.6cm]{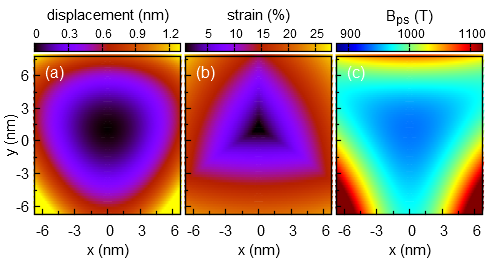}
  \caption{Contour plots of: (a) the displacement profile $|{\bf u({\bf r})}|$ of the triaxial strain;
  (b) strain distribution, (c) pseudo-magnetic field calculated using the the full value of the hopping
  parameter. The constant of the triaxial strain is: $c = 0.015 \text{ nm}^{-1}$. }
  \label{fig:tribump1}
\end{figure}
The pseudo-magnetic vector potential induced by strain in graphene
is given by Eq. (\ref{sc1}), and the pseudo-magnetic field is then
found as ${\bf B}_{ps} = {\nabla} \times {\bf A}_{ps}$. The vector
potential depends on the strained hopping parameter Eq.
(\ref{straint}), which can be expanded as,
\begin{equation}\label{strainexp}
 t_i/t_0 =  1 + \delta t^{(1)}_i + \delta t^{(2)}_i + \delta t^{(3)}_i \ldots,
\end{equation}
\begin{equation}
 t_i =  t_0 \left(1 - \beta \omega_i + \frac{1}{2} \beta^2 \omega_i^2 -
 \frac{1}{6} \beta^3 \omega_i^3 \ldots \right).
\end{equation}
Usually, only the first order term $\delta t^{(1)}_i$ is taken. Here, we will evaluate the effect of
the inclusion of the higher order terms.

A contour plot of the displacement profile of the triaxial strain is
shown in Figs. \ref{fig:tribump1}(a). The pseudo-magnetic field in
Fig. \ref{fig:tribump1}(c) is calculated using the full hopping
parameter from Eq. (\ref{straint}). The pseudo-magnetic field is
mostly homogeneous in the center. Away from the center, the
magnitude of the field follows the triangular shape of the
displacement with high magnitudes of the pseudo-magnetic field
corresponding to locations of large displacement.
\begin{figure}[h!]
  \centering
  \includegraphics[width=8.6cm]{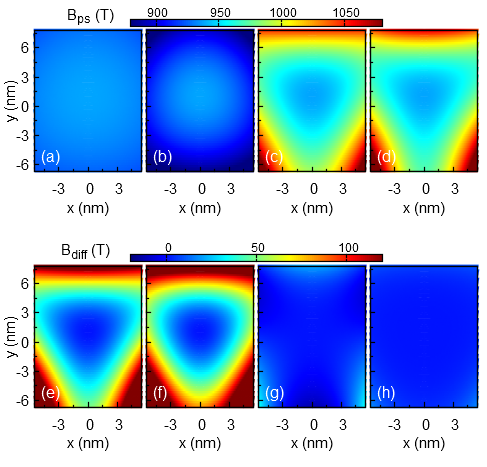}
  \caption{ Top: Contour plots of the pseudo-magnetic field generated from different approximations of
  the hopping parameter up to: (a) first, (b) second, (c) third and (d) fourth order. Bottom: (e,f,g,h)
  Difference plots between the respective approximations (a,b,c,d) and the field calculated using the
  full value of the hopping parameter, as in Fig. \ref{fig:tribump1}(d). The parameters of the triaxial
  strain is the same as in Fig. \ref{fig:tribump1}. }
  \label{fig:tribump2}
\end{figure}
In Fig. \ref{fig:tribump2} we plot the pseudo-magnetic field for
different approximations of the hopping parameter Eq.
(\ref{strainexp}). The figures are shown in pairs, with the top ones
presenting the magnitude of the field and the bottom ones presenting
the difference between the approximate and the full pseudo-magnetic
field calculated without approximations (see Fig.
\ref{fig:tribump1}(d)). Taking the first order approximation, see
Figs. \ref{fig:tribump2}(d) and (f), results in an almost completely
homogeneous pseudo-magnetic field, which shows large differences
compared to the full solution. Taking the second order approximation
results in a less homogeneous field which, however, shows a circular
symmetry instead of the triangular shape of the full field.
Calculating the pseudo-magnetic field using the third order
approximation finally shows the same triangular shape as the exact
pseudo-magnetic field. Adding the fourth order term further improves
the accuracy, but the correct shape has already been achieved with
the third order approximation.
\begin{figure}[h!]
  \centering
  \includegraphics[width=12cm]{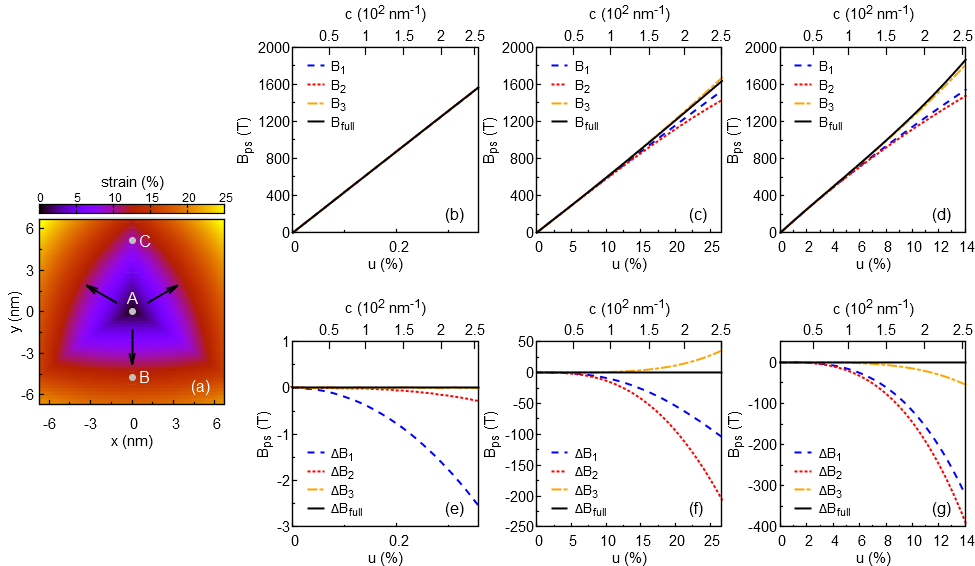}
  \caption{(a) Contour plot of the triaxial strain with three test points marked as A, B and C.
The arrows indicate the strain directions. The parameters of the
triaxial strain are the same as in Fig. \ref{fig:tribump1}. (b,c,d)
The pseudo-magnetic field as a function of the strain at the three
test points: (b) A, (c) B and (d) C. The field is calculated for
different approximations of the hopping parameter from first to
third order ($B_1$ to $B_3$ corresponds to $\delta t^{(1)}$ to
$\delta t^{(3)}$), as well as the full solution ($B_{full}$ for
$\delta t^{(full)}$). (e,f,g) The differences of the pseudo-magnetic
field approximations compared to the full expression ($\Delta B_i =
B_i - B_{full}$) at the three test points: (e) A, (f) B and (g) C.
In all cases the stain constant $c$ is scaled from $0$ to $0.025
\text{ nm}^{-1}$, as shown on the top $x$-axis. The resulting strain
at the test point is shown on the bottom $x$-axis.}
  \label{fig:ABC}
\end{figure}

In Fig. \ref{fig:ABC} we compare the pseudo-magnetic field
approximations at three points as function of applied strain. We
varied the $c$ parameter of the triaxial strain from $0$ to $0.025
\text{ nm}^{-1}$. The first point (A) is located in the center where
the strain remains very low (below $0.5\%$) even for high values of
$c$. Because of the low strain, all approximations are able to
accurately estimate the field.

Next, we considered point B, where the strain reaches up to $25\%$.
Because of the higher strain, the different approximations start to
diverge, although the differences aren't very large. The different
approximations diverge above $15\%$ strain but the differences
remain small even above $20\%$. However, point C shows more
significant differences. The approximations diverge at already for
$6\%$ strain. At high strain, the first order approximation
significantly underestimates the field (by as much as $350$ T).
Adding the second order term actually results in an even larger
underestimation of the field. Finally, adding the third order term
corrects the field magnitude so that it is in good agreement with
the full solution.

From these results we see that a correct estimation of the field in
point C is more difficult than in point B even though the maximum
strain is actually lower in point C. This is because the
pseudo-magnetic field depends not only on the intensity of the
strain, but also on the direction. The strain in point B is mostly
uniaxial, as it lies exactly along one of the three strain
directions (see Fig. \ref{fig:ABC}(a)). On the other hand, point B
feels a strong influence from both of the top strain directions, so
it is strongly non-uniaxial.
\section{Conclusions}
We investigated the pseudo magnetic field generated by
strain using the tight-binding approximation. The hopping
parameter and the deformation of the lattice vectors  are expanded up to
second order in the strain. The contribution of
the different terms are compared with the full numerical solution for
the pseudo magnetic field induced by a model triaxial strain.

For our numerical calculation a triaxial force is used to strain
graphene and we obtained the pseudo magnetic field resulting from
the different contributions resulting from different expansion terms
and compared the results with the full solution. Numerical results
for uniaxial strain clearly show that with applying strain the Fermi
velocity is spatial dependent. We included the second order term in
strain in the calculation of the pseudo magnetic field and showed
that the first order strain is reasonably valid up to $ 15\%$ strain
and that the pseudo magnetic field is the same in all ${\bf
K}$-points.
\section{Acknowledgment}
This work was supported by the Flemish Science Foundation (FWO-Vl),
the European Science Foundation (ESF) under the EUROCORES
Program EuroGRAPHENE within the project CONGRAN and the Methusalem programme of
the Flemish government.




\bibliographystyle{elsarticle-num}


\end{document}